\documentclass[12pt]{article}
\usepackage[a4paper,top=2.5cm,bottom=2.5cm,left=3.0cm,right=2.5cm]{geometry}

\usepackage{amsfonts}

\usepackage{amssymb}
\usepackage{makeidx}
\usepackage{amsmath}
\usepackage{graphicx}
\usepackage{epstopdf}

\begin{document}

\title{Nonlinear response of the quantum Hall system to a strong electromagnetic radiation}
\author{H. K. Avetissian, G. F. Mkrtchian \\
%EndAName
\small {Centre of Strong Fields Physics, Yerevan State University, Yerevan, Armenia}}

\maketitle

\begin{abstract}
We study nonlinear response of a quantum Hall system in semiconductor-hetero-structures via third harmonic generation process and nonlinear Faraday effect. We demonstrate that Faraday rotation angle and third harmonic radiation intensity have a characteristic Hall plateaus feature. These nonlinear effects remain robust against the significant broadening of Landau levels. We predict realization of an experiment through the observation of the third harmonic signal and Faraday rotation angle, which are within the experimental feasibility.
\end{abstract}

Integer quantum Hall effect (QHE) is remarkable phenomenon of two
dimensional electron gas (2DEG) systems, in which the longitudinal
resistance vanishes while the Hall resistance is quantized into plateaus 
\cite{QHE1}. The static QHE is the hallmark of dissipationless topological
quantum transport \cite{TKNN} and despite its long history\ there is a
continuing enormous amount of interest on this effect along various avenues.
With the advent of new materials, such as graphene and topological
insulators new regimes of QHE have been revealed \cite{Nov3,TI}. While
static properties of the integer QHE have been well investigated in the
scope of linear response theory, the dynamic and nonlinear responses in the
quantum Hall system (QHS) in the high-frequency regime are not fully
explored. In Ref. \cite{Kibis} considering the quantum dynamics of QHS
exposed to an intense high-frequency electromagnetic wave, it is shown that
the wave decreases the scattering-induced broadening of Landau levels.
Linear response of the QHS in the high-frequency regime has been
theoretically examined in Ref. \cite{MHA}. As was shown in Ref. \cite{MHA}
the plateau structure in the QHS is retained, up to significant degree of
disorder, even in the THz regime, although the heights of the plateaus are
no longer quantized. Then this effect has been confirmed experimentally in
Ref. \cite{MHA2}. Thus, a problem remains as how QHS responded to a strong
and high-frequency electromagnetic wave fields, which is the purpose of the
present study. In this case it is of interest to study generation of
harmonics \cite{Mer1, Mer2} at the interaction of a strong pump wave with
the Landau quantized 2DEG.

In the QHS wave-particle interaction can be characterized by the
dimensionless parameter $\chi =eE_{0}l_{B}/(\hbar \omega) $, which
represents the work of the wave electric field $E_{0}$ on the magnetic
length $l_{B}=\sqrt{c\hbar /(eB)}$ ($e$ is the elementary charge, $\hbar $
is Planck's constant, $c$ is the light speed in vacuum, and $B$ is the
magnetic field strength) in units of photon energy $\hbar \omega $. The
linear response theory is valid at $\chi <<1$. At $\chi \sim 1$ multiphoton
effects become considerable. In this paper we consider just multiphoton
interaction regime and look for features in the harmonic spectra of the
strong wave driven QHS. As a 2DEG system we consider GaAs/AlGaAs single
heterojunction. The time evolution of the considered system is found using a
nonperturbative numerical approach, revealing that the generated in the QHS
harmonics' radiation intensity has a characteristic Hall plateaus feature.
The effect remains robust against a significant broadening of Landau levels
and takes place for wide range of intensities and frequencies of a pump wave.

We begin our study with construction of the single-particle Hamiltonian
which defines the quantum dynamics of considered QHS. The 2DEG is taken in
the $xy$ plane ($z=0$) and a uniform static magnetic field is applied in the 
$\mathrm{OZ}$ direction. We consider an incoming electromagnetic radiation
pulse $E(t-z/c)$ propagating in the $\mathrm{OZ}$ direction and linearly
polarized along the x axis. The incoming wave is assumed to be
quasimonochromatic of carrier frequency $\omega $ and slowly varying
envelope $E_{0}(t)$. For the 2DEG as realized in GaAs/AlGaAs we have uniform
time-dependent electric field $E(t)=E_{0}(t)\sin \omega t$ and the
single-particle Hamiltonian of QHS reads: 
\begin{equation}
\mathcal{H}_{s}=\hbar \omega _{B}\left( \widehat{a}^{\dagger }\widehat{a}+%
\frac{1}{2}\right) +\left[ \frac{el_{B}E(t)}{\sqrt{2}}\left( \widehat{b}+i%
\widehat{a}\right) +\mathrm{h.c.}\right] .  \label{ham}
\end{equation}%
Here $\omega _{B}=eB/\left( m^{\ast }c\right) $ is the cyclotron frequency, $%
m^{\ast }=0.068m_{e}$ is the effective mass ($m_{e}$ - the bare electron
mass). For the interaction Hamiltonian we use a length gauge describing the
interaction by the potential energy. The ladder operators $\widehat{a}$ and $%
\widehat{a}^{\dagger }$ describe quantum cyclotron motion, while $\widehat{b}
$ and $\widehat{b}^{\dagger }$ correspond to guiding center motion. These
ladder operators satisfy the usual bosonic commutation relations $[\widehat{a%
},\widehat{a}^{\dagger }]=1$ and $[\widehat{b},\widehat{b}^{\dagger }]=1$.
The single free particle Hamiltonian, that is the first term in Eq. (\ref%
{ham}) can be diagonalized analytically. The wave function and energy
spectrum are given by: 
\begin{equation}
|\psi _{n,m}\rangle =|n,m\rangle ,  \label{WF}
\end{equation}%
\begin{equation}
\varepsilon _{n}=\hbar \omega _{B}\left( n+\frac{1}{2}\right) .
\label{energy}
\end{equation}%
Here $|n,m\rangle \ =|n\rangle \ \otimes |m\rangle $, with $|n\rangle $ and $%
|m\rangle $ being the harmonic oscillator wave functions. The eigenstates (%
\ref{WF}) are defined by the quantum numbers $n,m=0,1...$. Here $n$ is the
LL index. The LLs are degenerate upon second quantum number $m$ with the
degeneracy factor $N_{B}=\mathcal{S}/2\pi l_{B}^{2}$ which equals the number
of flux quanta threading the 2D surface $\mathcal{S}$ occupied by the 2DEG.
The terms $\sim \widehat{a}E(t)$ in the Hamiltonian (\ref{ham}) describe
transitions between LLs, while the terms $\sim \widehat{b}E(t)$ describe
transitions within the same LL. These transitions can be excluded from the
consideration by the appropriate dressed states for the construction of the
carrier quantum field operators. Expanding the fermionic field operator 
\begin{equation}
|\widehat{\Psi }\rangle =\sum\limits_{n,m}\widehat{a}_{n,m}|\widetilde{\psi }%
_{n,m}\rangle   \label{expand}
\end{equation}%
over the dressed states%
\begin{equation}
|\widetilde{\psi }_{n,m}\rangle =\exp \left[ -\frac{i}{\hbar }\frac{el_{B}}{%
\sqrt{2}}\int_{0}^{t}E(t^{\prime })dt^{\prime }\left( \widehat{b}^{\dagger }+%
\widehat{b}\right) \right] |\psi _{n,m}\rangle ,  \label{free}
\end{equation}%
the Hamiltonian of the system in the second quantization formalism 
\begin{equation*}
\widehat{H}=\left\langle \widehat{\Psi }\right\vert \mathcal{H}%
_{s}\left\vert \widehat{\Psi }\right\rangle 
\end{equation*}%
can be presented in the form:%
\begin{equation}
\widehat{H}=\sum\limits_{n=0}^{\infty }\sum\limits_{m=0}^{N_{B}}\varepsilon
_{n}\widehat{a}_{n,m}^{+}\widehat{a}_{n,m}+\sum\limits_{n,n^{\prime
}=0}^{\infty }\sum\limits_{m=0}^{N_{B}}E(t)\mathcal{D}_{n,n^{\prime }}%
\widehat{a}_{n,m}^{+}\widehat{a}_{n^{\prime },m},  \label{gr2}
\end{equation}%
where $\widehat{a}_{n,m}^{\dagger }$\ and $\widehat{a}_{n,m}$ are,
respectively, the creation and annihilation operators for a carrier in a LL
state, and $\mathcal{D}_{n,n^{\prime }}$ is the dipole moment operator: 
\begin{equation*}
\mathcal{D}_{n,n^{\prime }}=\frac{iel_{B}}{\sqrt{2}}\left[ \sqrt{n-1}\mathbb{%
\delta }_{n-1,n^{\prime }}+\sqrt{n}\mathbb{\delta }_{n,n^{\prime }-1}\right] 
\frac{\hbar \omega _{B}}{\varepsilon _{n^{\prime }}-\varepsilon _{n}}.
\end{equation*}%
Then we will pass to Heisenberg representation where operators obey the
evolution equation 
\begin{equation*}
i\hbar \frac{\partial \widehat{L}}{\partial t}=\left[ \widehat{L},\widehat{H}%
\right] 
\end{equation*}%
and expectation values are determined by the initial density matrix $%
\widehat{D}$: $<\widehat{L}>=Sp\left( \widehat{D}\widehat{L}\right) $. In
order to develop microscopic theory of the nonlinear interaction of the QHS
with a strong radiation field, we need to solve the Liouville-von Neumann
equation for the single-particle density matrix%
\begin{equation}
\rho (n_{1},m_{1};n_{2},m_{2},t)=<\widehat{a}_{n_{2},m_{2}}^{+}(t)\widehat{a}%
_{n_{1},m_{1}}(t)>  \label{grSPDM}
\end{equation}%
and for the initial state of the quasiparticles we assume an ideal Fermi gas
in equilibrium:%
\begin{equation}
\rho (n_{1},m_{1};n_{2},m_{2},0)=\frac{\delta _{n_{1},n_{2}}\delta
_{m_{1},m_{2}}}{1+\exp \left( \frac{\varepsilon _{n_{1}}-\varepsilon _{F}}{T}%
\right) }.  \label{grISPDM}
\end{equation}%
Including in Eq. (\ref{grISPDM}) quantity $\varepsilon _{F}$ is the Fermi
energy, $T$ is the temperature in energy units. As is seen from the
interaction term in the Hamiltonian (\ref{gr2}) quantum number $m$ is
conserved: $\rho (n_{1},m_{1};n_{2},m_{2},t)=\rho _{n_{1},n_{2}}\left(
t\right) \delta _{m_{1},m_{2}}$. To include the effect of the LLs broadening
we will assume homogeneous broadening of the LLs \cite{QHE2}. The latter can
be incorporated into evolution equation for $\rho _{n_{1},n_{2}}\left(
t\right) $ by the damping term $-i\Gamma _{n_{1},n_{2}}\rho
_{n_{1},n_{2}}\left( t\right) $ and from Heisenberg equation one can obtain
evolution equation for the reduced single-particle density matrix:%
\begin{equation}
i\hbar \frac{\partial \rho _{n_{1},n_{2}}(t)}{\partial t}=\left[ \varepsilon
_{n_{1}}-\varepsilon _{n_{2}}\right] \rho _{n_{1},n_{2}}(t)-i\Gamma
_{n_{1},n_{2}}\rho _{n_{1},n_{2}}\left( t\right)   \notag
\end{equation}%
\begin{equation}
-E(t)\sum\limits_{n}\left[ \mathcal{D}_{n,n_{2}}\rho _{n_{1},n}(t)-\mathcal{D%
}_{n_{1},n}\rho _{n,n_{2}}(t)\right] .  \label{grevol}
\end{equation}%
For the damping matrix we take $\Gamma _{n_{1},n_{2}}=\Gamma \left( 1-\delta
_{n_{1},n_{2}}\right) $, where $\Gamma $ measures the LL broadening.

Solving Eq. (\ref{grevol}) with the initial condition (\ref{grISPDM}) one
can reveal nonlinear response of the QHS to a strong radiation pulse. At
that one can expect intense radiation of harmonics of the incoming
wave-field in the result of the coherent transitions between LLs. The
harmonics will be described by the additional generated fields $%
E_{x,y}^{(g)} $. \ We assume that the generated fields are considerably
smaller than the incoming field $\left\vert E_{x,y}^{(g)}\right\vert
<<\left\vert E\right\vert $. In this case we do not need to solve
self-consistent Maxwell's wave equation with Heisenberg equations. To
determine the electromagnetic field of harmonics we can solve Maxwell's wave
equation in the propagation direction with the given source term:%
\begin{equation}
\frac{\partial ^{2}E_{x,y}^{(t)}}{\partial z^{2}}-\frac{1}{c^{2}}\frac{%
\partial ^{2}E_{x,y}^{(t)}}{\partial t^{2}}=\frac{4\pi }{c^{2}}\frac{%
\partial \mathcal{J}_{x,y}\left( t\right) }{\partial t}\delta \left(
z\right) .  \label{Max}
\end{equation}%
Here $\delta \left( z\right) $ is the Dirac delta function and $\mathcal{J}%
_{x,y}$ is the mean value of the surface current density operator: 
\begin{eqnarray}
\widehat{\mathcal{J}}_{x}\left( t\right) &=&\frac{-2e\hbar }{\sqrt{2}%
l_{B}m^{\ast }\mathcal{S}}\left\langle \widehat{\Psi }\right\vert \left( 
\hat{a}^{\dagger }+\hat{a}\right) \left\vert \widehat{\Psi }\right\rangle, 
\notag \\
\widehat{\mathcal{J}}_{y}\left( t\right) &=&\frac{-2e\hbar }{i\sqrt{2}%
l_{B}m^{\ast }\mathcal{S}}\left\langle \widehat{\Psi }\right\vert \left( 
\hat{a}^{\dagger }-\hat{a}\right) \left\vert \widehat{\Psi }\right\rangle.
\label{jqxy}
\end{eqnarray}%
With the help of Eqs. \ (\ref{expand}) and (\ref{grSPDM}) the expectation
value (\ref{jqxy}) of the total current in components can be written in the
following form:%
\begin{equation*}
\mathcal{J}_{x}\left( t\right) =j_{0}\sum\limits_{n=0}\sqrt{n+1}\mathrm{Re}%
\rho _{n,n+1}\left( t\right),
\end{equation*}%
\begin{equation}
\mathcal{J}_{y}\left( t\right) =-j_{0}\sum\limits_{n=0}\sqrt{n+1}\mathrm{Im}%
\rho _{n,n+1}\left( t\right) ,  \label{grcurry}
\end{equation}%
where $j_{0}=-\sqrt{2}e\hbar /\left( \pi l_{B}^{3}m^{\ast }\right) $ (here
we have taken into account\ the spin degeneracy factor). The solution to
equation (\ref{Max}) reads%
\begin{equation*}
E_{x,y}^{(t)}\left( t,z\right) =E_{x,y}\left( t-z/c\right)
\end{equation*}%
\begin{equation}
-\frac{2\pi }{c}\left[ \theta \left( z\right) \mathcal{J}_{x,y}\left(
t-z/c\right) +\theta \left( -z\right) \mathcal{J}_{x,y}\left( t+z/c\right) %
\right] ,  \label{sol}
\end{equation}%
where $\theta \left( z\right) $ is the Heaviside step function with $\theta
\left( z\right) =1$ for $z\geq 0$ and zero elsewhere. The first term in Eq. (%
\ref{sol}) is the incoming wave. In the second line of Eq. (\ref{sol}), we
see that after the encounter with the 2DEG two propagating waves are
generated. One traveling in the propagation direction of the incoming pulse
and one traveling in the opposite direction. The Heaviside functions ensure
that the generated light propagates from the source located at $z=0$. We
assume that the spectrum is measured at a fixed observation point in the
forward propagation direction. For the generated field at $z>0$ we have%
\begin{equation}
E_{x,y}^{(g)}\left( t-z/c\right) =-\frac{2\pi }{c}\mathcal{J}_{x,y}\left(
t-z/c\right).  \label{solut}
\end{equation}

Now, performing the summation in Eqs. (\ref{grcurry}) and using solutions (%
\ref{solut}) we can calculate the harmonic radiation spectrum with the help
of Fourier transform of the functions $E_{x,y}^{(g)}\left( t-z/c\right) $:%
\begin{equation}
E_{x,y}^{(g)}\left( s\right) =\frac{\omega }{2\pi }\int_{0}^{2\pi /\omega
}E_{x,y}^{(g)}\left( t\right) e^{is\omega t}dt.  \label{jxy}
\end{equation}

The spectrum contains in general both even and odd harmonics. However,
depending on the initial conditions, in particular, for the equilibrium
initial state (\ref{grISPDM}) the terms containing even harmonics cancel
each other because of inversion symmetry of the system and only the odd
harmonics are generated. The time evolution of system (\ref{grevol}) is
found with the help of the standard fourth-order Runge-Kutta algorithm and
for calculation of the power spectra the fast Fourier transform algorithm is
used. To avoid nonphysical effects semi-infinite pulses with smooth turn-on,
in particular, with hyperbolic tangent $tanh(t/\tau _{r})$ envelope is
considered. Here the characteristic rise time $\tau _{r}$ is chosen to be $%
\tau _{r}=10\pi /\omega $.

Figures 1 and 2 show nonlinear response of the QHS via normalized harmonics
field strengths versus Fermi energy for various pump wave intensities. Here
and below the temperature is taken to be $T/\hbar \omega _{B}=0.05$. Figure
1 displays normalized field strength at the fundamental harmonic $%
R_{y,1}=\left\vert E_{y}^{(g)}\left( 1\right) \right\vert /E_{0}$ polarized
perpendicular to the polarization of a pump wave, while Fig. 2 displays the
third harmonic field strength $R_{x,3}=\left\vert E_{x}^{(g)}\left( 3\right)
\right\vert /E_{0}$. From these figures we immediately notice a step-like
structure of the nonlinear response of the QHS system as a function of $%
\varepsilon _{F}$ for various pump wave intensities. Although the step
heights are not quantized exactly, the flatness, which is a intrinsic
property of the static QHE, surprisingly exists also in the nonlinear
response of the QHS. In the static QHE the step structure of the Hall
conductivity is a quantum and topological effect. In the considered case
Eqs. (\ref{grcurry}) does not simply reduce to a topological expression and
the result for the robust plateaus of the nonlinear optical response is not
apparent.

We further examine how the step-like structure in the nonlinear response of
the QHS behaves for various pump wave frequencies. The generated fields
versus Fermi energy and pump wave frequency at the fundamental and third
harmonics are shown in Figs. 3 and 4. Thus, the step structure preserves for
the wide range of the pump wave frequencies. 

\begin{figure}[tbp]
\includegraphics[width=.5\textwidth]{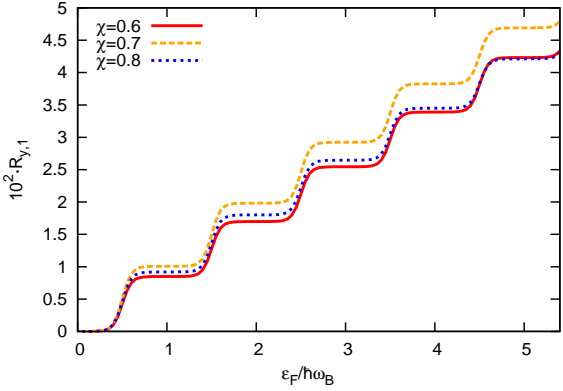}
\caption{Nonlinear response of QHS system via normalized field strength
versus Fermi energy at the fundamental harmonic polarized perpendicular to
the incoming wave for various intensities with $\protect\omega _{B}=1.5%
\protect\omega $. The LL broadening is taken to be $\Gamma =0.1\hbar \protect%
\omega _{B}$. }
\label{2w}
\end{figure}

\begin{figure}[tbp]
\includegraphics[width=.5\textwidth]{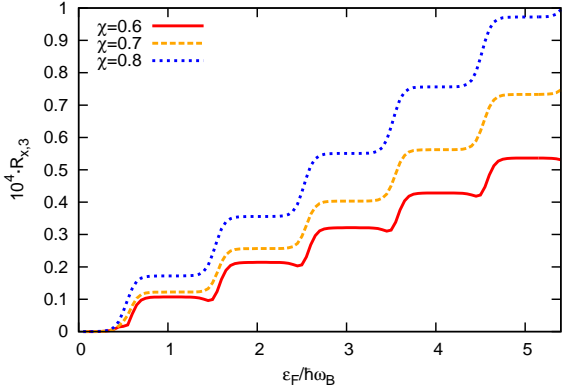}
\caption{The third harmonic normalized field strength in the QHS versus
Fermi energy for various pump wave intensities with $\protect\omega _{B}=1.5%
\protect\omega $. The LL broadening is taken to be $\Gamma =0.1\hbar \protect%
\omega _{B}$. }
\label{eps3}
\end{figure}

\begin{figure}[tbp]
\includegraphics[width=.5\textwidth]{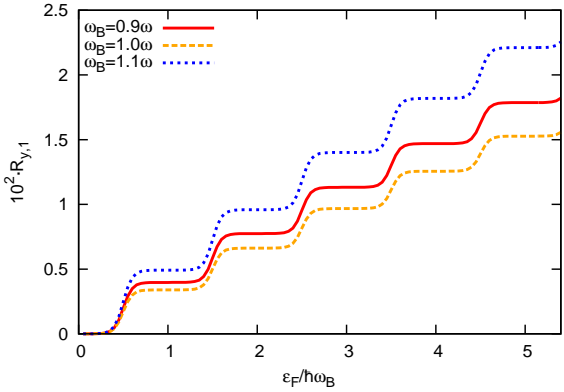}
\caption{Nonlinear response of QHS system via normalized field strength
versus Fermi energy at the fundamental harmonic polarized perpendicular to
the incoming wave for various wave frequencies with $\protect\chi =0.7$. The
LL broadening is taken to be $\Gamma =0.1\hbar \protect\omega _{B}$. }
\end{figure}
\begin{figure}[tbp]
\includegraphics[width=.5\textwidth]{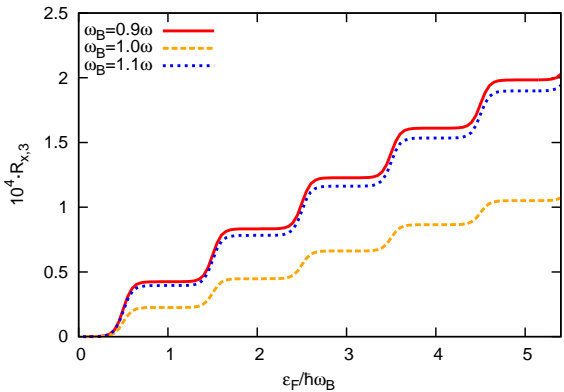}
\caption{The third harmonic normalized field strength in the QHS versus
Fermi energy for various pump wave frequencies with $\protect\chi =0.8$. The
LL broadening is taken to be $\Gamma =0.1\hbar \protect\omega _{B}$. }
\end{figure}

We also investigate how the step-like structure in the nonlinear response of
the QHS behaves as we vary the LL broadening. So we have calculated $R_{y,1}$
as a function of $\Gamma $, for fixed values of $\omega $ and $\chi $. We
can see from Fig. 5 that, while the density of states broadens with a width $%
\sim \Gamma $ the step structure remains up to large $\Gamma $.

\begin{figure}[tbp]
\includegraphics[width=.5\textwidth]{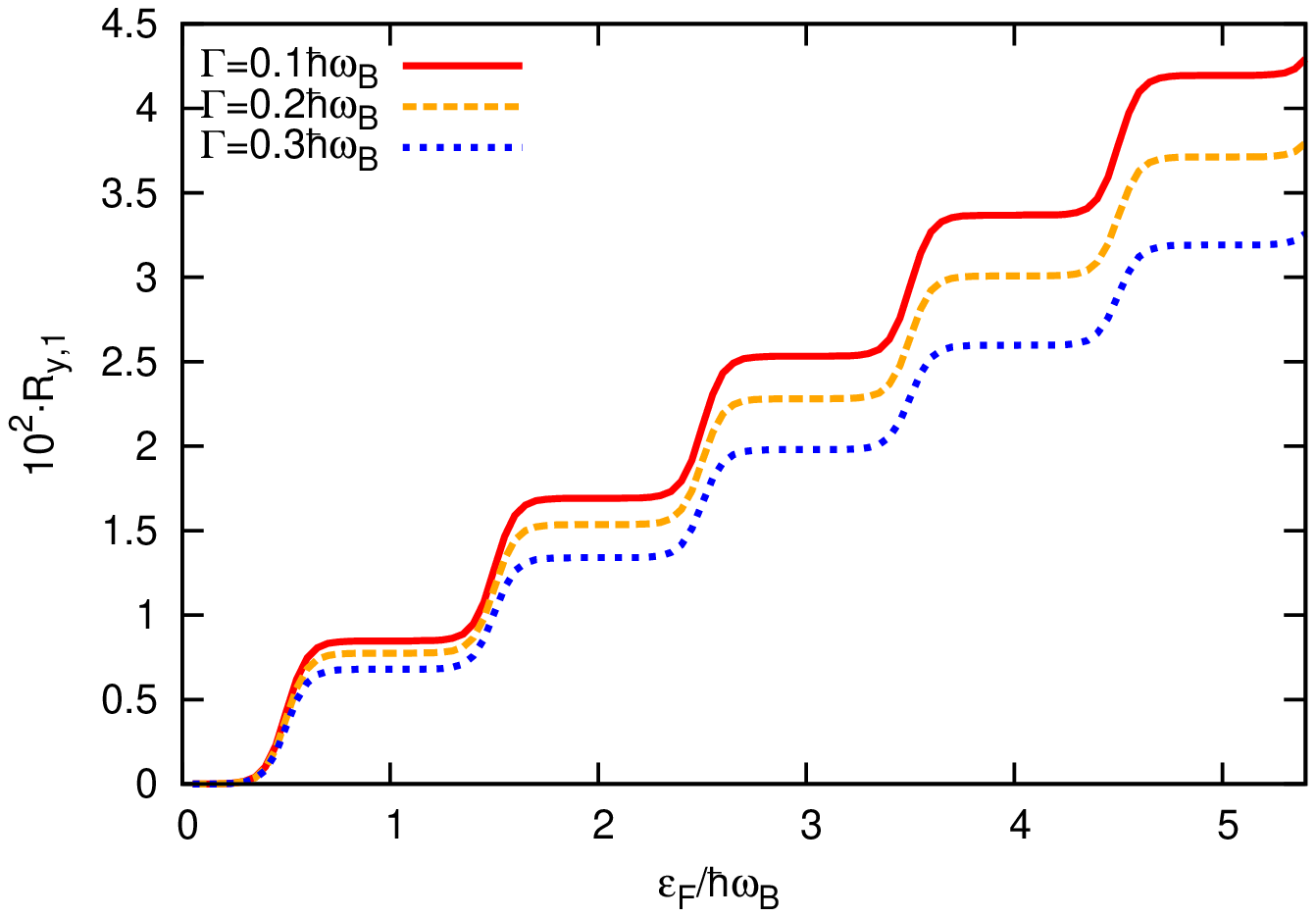}
\caption{The fundamental harmonic normalized field strength in the QHS
versus Fermi energy for various values of LL broadening at the fixed pump
wave frequency $\protect\omega _{B}=2\protect\omega $. The intensity
parameter is taken to be $\protect\chi =0.7$.}
\label{eps6}
\end{figure}

Finally let us consider the experimental feasibility. It is clear that in
experiment one can observe the considered effect by measuring $R_{y,1}$
and/or $R_{x,3}$. The first quantity is responsible for the nonlinear
Faraday effect, while last quantity responsible for third harmonic radiation
polarized along the incoming wave polarization. Thus, the step structure
should be observed as jumps in the intensity of third harmonic or
fundamental harmonic radiation with orthogonal polarization. The magnetic
field strength is assumed to be $B=3\ T$. For the incoming wave field we
will assume $\hbar \omega \simeq 3.5\ \mathrm{meV}$. The intensity of the
incoming wave for $\chi =0.8$ is $4.\,\allowbreak 4\times 10^{3}$ $\mathrm{%
W/cm}^{2}$. For the setup of Fig. 1 the steps in the Faraday-rotation angle $%
\Delta \Theta _{F}\sim R_{y,1}\sim 10\ \mathrm{mrad}$\textrm{, \ }which is
well within the experimental resolution \cite{Exp}. For the setup of Fig. 2
with the chosen parameters the average intensity of the third harmonic
radiation is $I_{3}\sim 2.\,\allowbreak 2\times 10^{-5}$ $\mathrm{W/cm}^{2}$
(corresponding to $10^{16}$ $\mathrm{photons/s\cdot cm}^{2}$) with the steps 
$\Delta I_{3}\sim 0.36\times I_{3}$.

To summarize, we have presented a microscopic theory of the 2DEG interaction
with coherent electromagnetic radiation in the quantum Hall regime. The
evolutionary equation for a single-particle density matrix has been solved
numerically. We have revealed that the nonlinear optical response of QHS to
an intense radiation pulse, in particular, radiation intensity at the
harmonics, as well as nonlinear Faraday effect, has a characteristic Hall
plateau structures that persist for a wide range of the pump wave
frequencies and intensities even for significant broadening of \ LLs.

This work was supported by the RA MES State Committee of Science, in the
frames of the research project No. 15T-1C013.

\end{document}